\documentclass[apj]{emulateapj}
\usepackage{natbib}
\usepackage{graphicx}
\usepackage[toc,page]{appendix}
\usepackage{natbib}
\usepackage{color}
\usepackage{threeparttable}
\usepackage{hyperref}
\usepackage{breakurl}
\begin{document}

\slugcomment{Accepted for publication in the Astrophysical Journal Letters}
\title{Quiescent Galaxies in the 3D-HST Survey: Spectroscopic confirmation of a large number of galaxies with relatively old stellar populations at $z\sim2$}
\email{kate.whitaker@nasa.gov}
\author{Katherine E. Whitaker\altaffilmark{1,8}, Pieter G. van Dokkum\altaffilmark{2}, 
Gabriel Brammer\altaffilmark{3}, Ivelina G. Momcheva\altaffilmark{2}, Rosalind Skelton\altaffilmark{2},
Marijn Franx\altaffilmark{4}, Mariska Kriek\altaffilmark{5}, Ivo Labb\'{e}\altaffilmark{4},
Mattia Fumagalli\altaffilmark{4}, Britt F. Lundgren\altaffilmark{6},
Erica J. Nelson\altaffilmark{2}, Shannon G. Patel\altaffilmark{4}, Hans-Walter Rix\altaffilmark{7}}
\altaffiltext{1}{Astrophysics Science Division, Goddard Space Flight Center, Code 665, Greenbelt, MD 20771, USA}
\altaffiltext{2}{Department of Astronomy, Yale University, New Haven, CT 06520, USA}
\altaffiltext{3}{European Southern Observatory, Alonso de C\'{o}rdova 3107, Casilla 19001, Vitacura, Santiago, Chile}
\altaffiltext{5}{Department of Astronomy, University of California, Berkeley, CA 94720, USA}
\altaffiltext{4}{Sterrewacht Leiden, Leiden University, NL-2300 RA Leiden, The Netherlands}
\altaffiltext{6}{Department of Astronomy, University of Wisconsin, Madison, WI 53706, USA}
\altaffiltext{7}{Max Planck Institut fur Astronomie, K\"{o}nigstuhl 17, D-69117 Heidelberg, Germany}
\altaffiltext{8}{NASA Postdoctoral Program Fellow}
\shortauthors{Whitaker et al.}
\shorttitle{Spectroscopic confirmation of quiescent galaxies at $z\sim2$}

\begin{abstract}
Quiescent galaxies at $z\sim2$ have been identified in large numbers based on rest-frame colors, but
only a small number of these galaxies have been spectroscopically confirmed to show that their rest-
frame optical spectra show either strong Balmer or metal absorption lines. Here, we median stack
the rest-frame optical spectra for 171 photometrically-quiescent galaxies at $1.4<z<2.2$ from the
3D-HST grism survey. In addition to H$\beta$ ($\lambda$4861$\mathrm{\AA}$), we unambiguously identify 
metal absorption lines in the stacked spectrum, including the G-band ($\lambda$4304$\mathrm{\AA}$), 
Mg I ($\lambda$5175$\mathrm{\AA}$), and Na I ($\lambda$5894$\mathrm{\AA}$).
This finding demonstrates that galaxies with relatively old stellar populations already existed when the universe 
was $\sim3$ Gyr old, and that rest-frame color selection techniques can efficiently select them.
We find an average age of $1.3^{+0.1}_{-0.3}$ Gyr when fitting a simple stellar population to the
entire stack. We confirm our previous result from medium-band photometry that
the stellar age varies with the colors of quiescent galaxies: the reddest 80\%
of galaxies are dominated by metal lines and have a relatively old mean age of
$1.6^{+0.5}_{-0.4}$ Gyr, whereas the bluest (and brightest) galaxies have strong Balmer
lines and a spectroscopic age of $0.9^{+0.2}_{-0.1}$ Gyr.
Although the spectrum is dominated by an evolved stellar population, we also find [OIII] and H$\beta$ emission.
Interestingly, this emission is more centrally
concentrated than the continuum with $L_{\mathrm{OIII}}=1.7\pm0.3\times10^{40}$ erg s$^{-1}$, 
indicating residual central star formation or nuclear activity.  
\end{abstract}


\section{Introduction}
\label{sec:intro}

In the nearby universe, the most massive galaxies almost always have little ongoing star-formation and red colors that
exhibit a remarkably small intrinsic scatter \citep[e.g.,][]{Bower92}.  These red, quiescent galaxies form a well-defined
color-mass relation known as the ``red sequence''.  Photometric studies of large, representative samples of galaxies 
based on broad-band and medium-band photometry have pushed the detection of the red sequence out to $z\sim2$ 
\citep[e.g.,][]{Williams09,Whitaker11,Brammer11, Nicol11}.  However, only a small fraction of
these distant quiescent galaxies have been spectroscopically 
confirmed \citep[e.g.,][]{Cimatti04, Daddi05, Kriek06, Cimatti08, Kriek08b, Kriek09a, vanDokkumBrammer10, vandeSande11, vandeSande12, Onodera12, Toft12, Bezanson13}.

Interestingly, most distant quiescent galaxies with high quality rest-frame optical spectra appear to exhibit young ages, showing 
strong Balmer absorption lines \citep[e.g.,][]{vandeSande12, Bezanson13}.  This could imply that the bulk of high
redshift quiescent galaxies were recently quenched.
However, most spectroscopic studies in the literature are biased                            
towards the brightest (and consequently youngest) of such galaxies.  
Therefore, spectroscopic ages for a representative sample
of quiescent galaxies remains elusive.

Owing to the near-infrared (NIR) slitless spectroscopic capabilities provided
by the Wide-Field Camera 3 (WFC3) on the 
Hubble Space Telescope (HST), it is now possible to obtain
low-resolution spectroscopy of a mass-limited sample
of quiescent galaxies at $z\sim2$.  In this letter, we stack the spectra of 
171 quiescent galaxies at $z\sim2$ and demonstrate that they have absorption features indicative of evolved
stellar populations.  
Furthermore, with ages derived from the stacked grism spectra, we are in a unique position to 
test what drives the color spread of quiescent galaxies for the first time.

We assume a $\Lambda$CDM cosmology with $\Omega_{M}$=0.3, $\Omega_{\Lambda}$=0.7, 
and $H_{0}$=70 km s$^{-1}$ Mpc$^{-1}$ throughout the paper.  All magnitudes are 
given in the AB system.

\section{Data}
\label{sec:data}

The 3D-HST treasury program \citep{Brammer12}, a 248--orbit NIR
spectroscopic survey with the HST/WFC3 G141 grism, provides spatially resolved low-resolution spectra of all
objects in five well-studied extra-galactic fields to a $5\sigma$ depth for continuum magnitudes 
of $H_{\mathrm{F140W}}\sim23$.
3D-HST has targeted the AEGIS, COSMOS, GOODS-S and UDS fields, as well as incorporated
publicly-available data in the GOODS-N field (GO:11600; PI:Weiner).  WFC3 imaging for 
all five fields is available from the CANDELS survey~\citep{Grogin11,Koekemoer11}.
With a wavelength range of 1.10$\mu$m$ < \lambda < $1.65$\mu$m,
the prominent age-dependent H$\beta\lambda$4861 and Mg$\lambda$5175 absorption features in quiescent galaxies at $z\sim2$ are
observable.  The first-order dispersion of the G141 grism is 46 $\mathrm{\AA}$/pixel ($\mathrm{R}\sim130$) with a spatial resolution of 
0.$^{\prime\prime}$12, sampled with 0.$^{\prime\prime}$06 pixels; as the spectra have high spatial resolution 
and low spectral resolution, the line width almost exclusively reflects the size 
of the galaxy in the dispersion direction.

The sample used in this letter is selected from WFC3-selected photometric catalogs generated 
from the \textit{HST} ACS and WFC3 images of the CANDELS and 3D-HST survey fields, along 
with ancillary ground-based optical and NIR images and mid-IR images with \textit{Spitzer}/IRAC 
using a methodology similar to that described by \citet{Whitaker11}.  A full description of 
these catalogs is beyond the scope of this letter and will be described in a forthcoming paper 
(Skelton et al. in prep).  We extract a two-dimensional flux-calibrated WFC3/G141 grism 
spectrum for every object in the photometric catalog with $H_{\mathrm{F140W}}<24$.  Details 
of the reduction and extraction of the grism spectra, including accounting for the 
contamination of overlapping objects, are provided by \citet{Brammer12,Brammer12b}.

To determine the galaxy redshifts, we first compute a purely photometric redshift from the photometry,
using the EAZY code \citep{Brammer08}.  We then fit the full two-dimensional grism spectrum 
separately with a combination of the continuum template taken from the EAZY fit and a 
single emission-line-only template with fixed line ratios taken from the SDSS composite 
star-forming galaxy spectrum of \citet{Dobos12}.  The final grism redshift, {\tt z\_grism}, 
is determined on a finely-sampled redshift grid with the photometry-only redshift probability 
distribution function used as a prior.  This method is somewhat more flexible than that 
originally described by \citet{Brammer12}, but the redshift precision is similar with 
$\sigma\sim 0.0035(1+z)$.  Finally, rest-frame colors and stellar population parameters are 
computed from the photometry with the EAZY and FAST \citep{Kriek09a} codes, respectively, 
with the redshift fixed to {\tt z\_grism} and assuming a \citet{Chabrier} initial mass function.

\section{Sample Selection}
\label{sec:selection}

A standard method for discriminating high redshift quiescent galaxies from star-forming galaxies is selecting on the
rest-frame $U$--$V$ and $V$--$J$ 
colors \citep[e.g.,][]{Labbe05, Wuyts07, Williams09,Bundy10,Cardamone10b,Whitaker11,Brammer11,Patel12}; 
quiescent galaxies have strong Balmer/4000\AA\ breaks, characterized by red $U$--$V$ colors and relatively 
blue $V$--$J$ colors.

\begin{figure}[t]
\leavevmode
\centering
\includegraphics[width=0.9\linewidth]{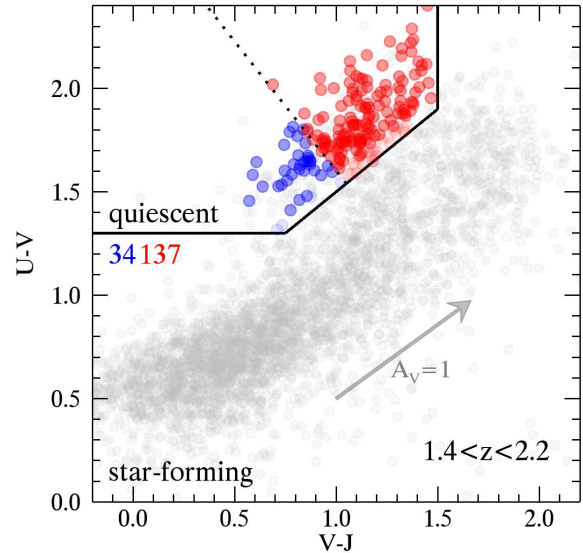}
\caption{Rest-frame $U$--$V$ and $V$--$J$ color selection of quiescent 
galaxies, further separating the sample into two bins of younger (blue) and older (red) quiescent galaxies in Section~\ref{sec:colors}. All galaxies with $H_{\mathrm{F140W}}<24$ are shown in grayscale.}
\label{fig:uvj}
\end{figure}

Following the definition of \citet{Whitaker12a}, our quiescent selection box is shown in Figure~\ref{fig:uvj},
with the larger 3D-HST parent sample at $1.4<z<2.2$ shown in grey-scale.  
Using the dotted line in Figure~\ref{fig:uvj} ($(U-V)=-1.25\times(V-J)+2.85$), we further divide our 
quiescent sample into ``younger'' (blue) and ``older'' (red) galaxies (Section~\ref{sec:colors}).
The nature of the stellar populations of galaxies
with rest-frame colors close to the quiescent/star-forming division is not clear and we
therefore restrict our analysis to be more conservative by 0.08 mag (excluding those galaxies with high
transparency in Figure~\ref{fig:uvj}).
We select galaxies at $1.4<z_{\mathrm{grism}}<2.2$ with stellar masses log(M$_{\star})>10.5$ M$_{\odot}$, further restricting our sample
to require $>75\%$ wavelength coverage of the grism spectrum with $<50\%$ of the pixels flagged as bad.
The grism spectra of quiescent galaxies do not have strong emission line features and therefore the redshifts may not be reliable beyond certain 
magnitude limits; we additionally require that we detect the continuum at $\gtrsim5\sigma$ per resolution element in the 1D spectrum ($H_{\mathrm{F140W}}<22.8$), removing the faintest 20\% of galaxies.

\section{Quiescent Stacks}
\label{sec:stacks}

Although the individual galaxies are too faint to discern spectral features, by
stacking the spectra of our final sample of 171 galaxies we can achieve the necessary signal-to-noise ratio
to robustly identify absorption features in a large well-defined high redshift sample for the first time.
To normalize the spectra, we shift to the rest-frame and interpolate the flux values 
to a new wavelength grid with 10$\mathrm{\AA}$ width bins.
Next, we fit the continuum in each spectrum with a third order polynomial, masking regions around prominent absorption features.  
After dividing by the continuum, we determine the median flux value at each wavelength bin and smooth with a 
boxcar of 20$\mathrm{\AA}$. The grey-scale
error bars are derived from 100 bootstrap iterations of the stacking analysis.

\begin{figure*}[t!]
\leavevmode
\centering
\includegraphics[width=0.75\linewidth]{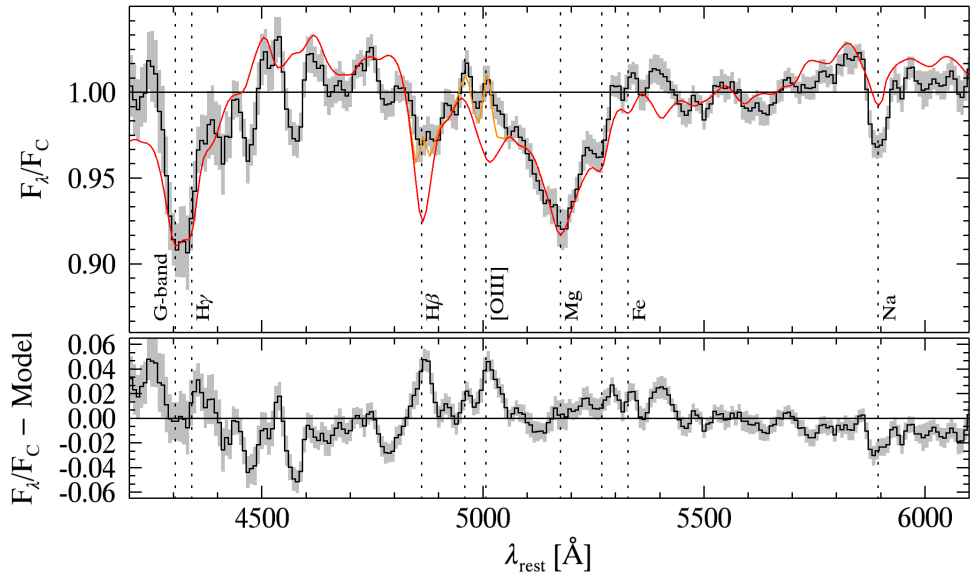}
\caption{Median rest-frame stack of 171 quiescent galaxies (top) and the residuals after subtracting the best-fit model (red) with 
an age of 1.25 Gyr (bottom), using the grism redshifts.  The orange model shows an additional 5\% enhancement
of centrally-concentrated H$\beta$ and [OIII] emission. 
The accuracy of the grism redshifts enables us to resolve absorption features at $z\sim2$ with high signal-to-noise for the first time.}
\label{fig:stacks_zphot}
\end{figure*}

We show the stacked spectrum in Figure~\ref{fig:stacks_zphot}.  In the top panel, 
we see many clear absorption features: G-band ($\lambda4304\mathrm{\AA}$)
blended with H$\gamma$($\lambda4341\mathrm{\AA}$), Mg I ($\lambda5175\mathrm{\AA}$), Fe I ($\lambda5269\mathrm{\AA}$),
and Na I ($\lambda5894\mathrm{\AA}$).  These metal lines are indicative of an evolved stellar population.
We caution that systematic residuals in the blue part of the spectrum are of similar
strength as the emission lines. The residuals do not seem to be caused by photon noise,
and show up in all stacked spectra.
We have not identified the cause of these residuals.
We cannot exclude errors in the stellar population synthesis models, although that is
unlikely in this well-studied wavelength range.

We perform a least-squares minimization using the \citet{Vazdekis10} models to find the best-fit age of the stack.
The \citet{Vazdekis10} models provide moderately high resolution SEDs computed assuming an
instananeous burst, Solar metallicity and a \citet{Kroupa01} universal initial mass function,
based on the MILES \citep{SanchezBlazquez06} stellar library.
We simulate the expected morphological broadening
by convolving the high-resolution models with the grism response and the object morphology in the 
$H_{\mathrm{F140W}}$ direct image \citep[see][for more information on how the 2-D model spectra are generated]{Brammer12b}.
One-dimensional model spectra are then extracted in the same way as was done for the observed G141 spectra.
The red line in Figure~\ref{fig:stacks_zphot} is the best-fit
convolved model with an age of 1.25 Gyr, fit to all wavelengths.

\begin{figure}[t!]
\leavevmode
\centering
\includegraphics[width=0.8\linewidth]{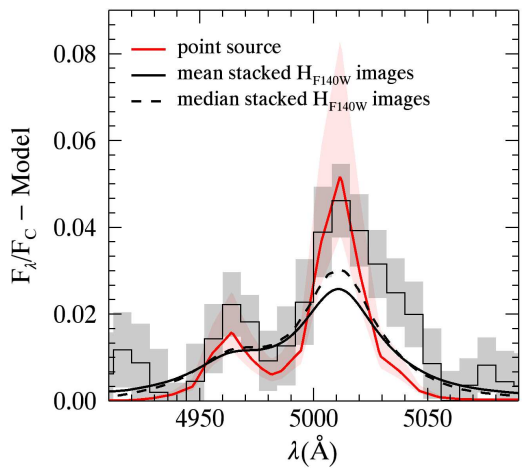}
\caption{The morphology of the residual [OIII] emission after subtracting a 1.25 Gyr model (histogram).   The morphology is compared to the expectation for a centrally-concentrated gas distribution (red line) and to a distribution that follows the galaxy light (black line).  
The red shaded region indicates the $1\sigma$ confidence interval resulting from the modeling analysis.  
The resolved double-peak of the [OIII] feature is only expected for centrally-concentrated emission 
with the low grism resolution.}
\label{fig:O3}
\end{figure}

The bottom panel of Figure~\ref{fig:stacks_zphot} shows the residuals when subtracting the best-fit model (red).
Here, we see the [OIII] doublet ($\lambda4959, 5007\mathrm{\AA}$) and H$\beta$ in emission.
It is somewhat surprising to clearly resolve the [OIII] doublet given the morphological broadening in the grism spectra.  
In Figure~\ref{fig:O3}, we demonstrate the expected resolution for the [OIII] emission feature at a fixed
line ratio of 3:1 when 
convolving with the interlaced PSF (red) and the average galaxy $H_{\mathrm{F140W}}$ profile 
(black).  The stacked profile is constructed by re-centering, masking neighbors 
and normalizing by the total flux of individual galaxies before taking the mean/median.  
The [OIII] doublet
is blended together when the emission is a global feature of the galaxy, whereas centrally-concentrated emission (within roughly the central
pixel) shows two clear peaks.   The point source nature of the [OIII] emission lines also indicates that the high quality of the grism
redshifts, as redshift errors would broaden the observed [OIII] lines in the stack.
We note that although the relative ratio of the [OIII] doublet is mildly 
sensitive to the assumed best-fit model, this 
effect is significantly less than the error bars from the stacking analysis themselves.  
The residuals are consistent with both centrally-concentrated [OIII] and H$\beta$ emission.

To model the central H$\beta$ and [OIII] enhancements, 
we simultaneously fit a grid of \citet{Vazdekis10} models convolved with the average galaxy profile with enhancements
ranging from 0--15\% above the continuum flux for H$\beta$ and [OIII] emission 
lines convolved with the PSF.
We assume a fixed line ratio of $\mathrm{[OIII]}\lambda5007/\mathrm{H}\beta=1$, 
consistent with the residuals in Figure~\ref{fig:stacks_zphot} and 
line ratio diagnostics for post-starburst galaxies in the Sloan Digital Sky Survey \citep[SDSS;][]{Mendel13}.
The best-fit model of 1.25 Gyr with a 5\% [OIII] and H$\beta$ central enhancement is shown in the top panel
of Figure~\ref{fig:stacks_zphot} (orange).  

We measure the [OIII]$\lambda5007$ line flux from the residuals in Figure~\ref{fig:stacks_zphot} to be $1.4\pm0.2\times10^{-18}$ 
erg s$^{-1}$ cm$^{-2}$, which corresponds to $L_{\mathrm{OIII}}$ of 
$1.7\pm0.3\times10^{40}$ erg s$^{-1}$ for the median redshift of 1.64.
Similarly, we measure the H$\beta$ line flux to be $1.5\pm0.2\times10^{-18}$ erg s$^{-1}$ cm$^{-2}$.  
To determine the line flux errors, we add the simulation errors
in quadrature to the error in the average continuum level.
The resulting [OIII]$\lambda5007$/H$\beta$ ratio is $0.9\pm0.2$.  
If we further correct $L_{\mathrm{OIII}}$ for the average dust extinction value of $A_{V}=0.5$ determined from the photometry, 
we estimate the dust-corrected [OIII] luminosity to 
be $L^{c}_{\mathrm{OIII}}=2.4\times10^{40}$ erg s$^{-1}$,
following \citet{Bassani99}.  Adopting the conversion from $L^{c}_{\mathrm{OIII}}$
to $L_{\mathrm{X}}$ from \citet{Lamastra09} and a factor of ten conversion from $L_{\mathrm{X}}$
to $L_{bol}$ \citep{Lusso10}, we estimate a typical active galactic nucleus (AGN) luminosity to be $L_{bol}=2\times10^{42}$ erg s$^{-1}$.
However, we note that it is not possible to differentiate between centrally-concentrated residual star-formation
or an AGN without further information. 

\section{What drives the spread in rest-frame colors?}
\label{sec:colors}

\begin{figure*}[t]
\leavevmode
\centering
\includegraphics[width=0.85\linewidth]{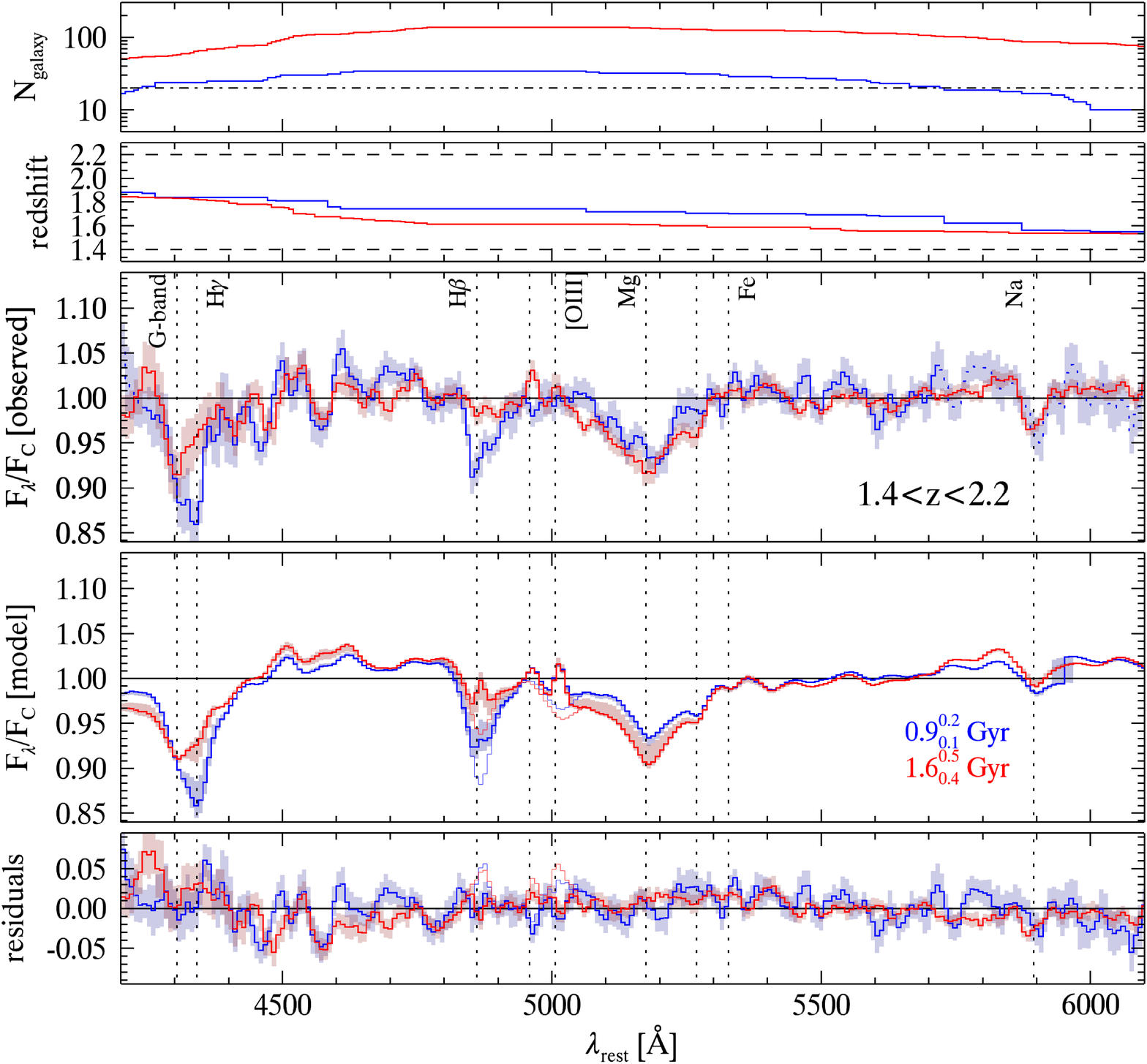}
\caption{Median stacks of quiescent galaxies in two rest-frame color bins (as defined in Figure~\ref{fig:uvj}).  The
top two panels show the number of galaxies and median redshift of each wavelength bin.  The observed stacks
in the third panel reveal that the bluest quiescent galaxies have younger
ages, most notable by the stronger H$\beta$ and weaker Mg I absorption.  The fourth panel shows the best-fit models, 
including a 5--6\% centrally-concentrated enhancement of H$\beta$ and [OIII] emission. The 
residuals are shown in the bottom panel.}
\label{fig:stacks}
\end{figure*}

With these high quality stacked spectra, we can now begin to assess the driving factor behind the 
spread in rest-frame colors of quiescent galaxies.  The three possibilities include age, dust or metallicity variations
amongst galaxy populations.  The difference in rest-frame $U$--$V$ and $V$--$J$ colors corresponding to 1 Gyr of passive aging 
within the \citet{BC03} models results in a 0.4 mag vector parallel to the quiescent sequence.  
Similarly, increasing the metallicity from 
log(Z)=0.02 (Solar) to log(Z)=0.05 or adding 0.5 mag of dust reddening following the \citet{Calzetti00} dust law 
results in an almost identical vector in both direction and magnitude.  Due to degeneracies
between age, dust and metallicity, it is notoriously difficult to differentiate which of these effects dominates.  

Metallicity variations will cause the color spread in galaxies to increase over time as the galaxies passively evolve, reaching
a roughly constant scatter for the oldest stellar populations.
As noted by \citet{Whitaker10}, the trend for the intrinsic scatter in the rest-frame colors of massive quiescent
galaxies to steadily decrease from $z=2$ to the present epoch  \citep[see also, e.g.,][]{Ruhland09, Papovich10}
is opposite to that predicted from metallicity variations alone.
We therefore expect that metallicity will not have a large effect at high redshifts.

As the shape of the Balmer/4000\AA\ break depends sensitively on the treatment of dust reddening, one would 
ideally need to correlate direct, independent measurements of the ages of massive galaxies with 
their colors.
Surprisingly, \citet{Bezanson13} did not find a simple
correlation between the rest-frame color and the strength of the Balmer absorption lines, as some of the reddest
galaxies in the spectroscopic sample had strong Balmer absorption lines.  However, with a small
sample size of 13 galaxies, it is difficult to gauge
if these results are representative of the overall quiescent population.

In Figure~\ref{fig:stacks}, we plot the median stacked spectra in two rest-frame
color bins for the 171 quiescent galaxies, as color-coded in Figure~\ref{fig:uvj}.  
The total number of galaxies in each bin is labeled in Figure~\ref{fig:uvj} 
and the number of galaxies per wavelength bin is indicated in the top histogram
of Figure~\ref{fig:stacks}.  A minimum number of 20 galaxies was required per 
wavelength bin to ensure a robust measurement.

Here, we assess the effects of age as a function of rest-frame color for a 
mass-limited sample of quiescent galaxies
by measuring the strength of absorption features such as Mg I and H$\beta$.
Both of these spectral features are sensitive to the age of the stellar population, 
but not very sensitive to metallicity or dust reddening.  Given the narrow baseline of the spectral features,
dust extinction is not expected to have a significant effect.
At the grism resolution, the H$\beta$ absorption feature is 
expected to be about 4--5\% stronger in a 1 Gyr stellar population as compared to a 2 Gyr stellar population, whereas the 
Mg I absorption feature is about 3--4\% weaker.  Increasing the metallicity results in a slightly deeper Mg I feature, but only 
on the order of $\lesssim1\%$.  To first order, any change measured in H$\beta$ and Mg I 
is due to age differences between the galaxies.  

The average age of galaxies with the bluest
rest-frame colors is $0.9^{+0.2}_{-0.1}$ Gyr, consistent with the expectation that they are in
the ``post-starburst'' phase \citep{Kriek10, Whitaker10, Whitaker12a}.  These galaxies clearly
show significantly stronger H$\beta$ and H$\gamma$ absorption than the reddest galaxies.  Note 
the difference in shape of the blended G-band and H$\gamma$ between these two stacks.  The redder galaxies
are 0.7 Gyr older on average, with a best-fit age of $1.6^{+0.5}_{-0.4}$ Gyr.  
For a 0.7 Gyr age difference, the \citet{BC03} models predict $\Delta(U-V)=0.2$ and 
$\Delta(V-J)=0.3$, corresponding to exactly the observed difference in the median rest-frame
colors of the two bins.

The older quiescent galaxies show clear signs of either central residual star-formation activity or an AGN, 
while the enhancement in the younger
galaxies is not well-constrained despite a well-determined spectroscopic age.
We measure the residual [OIII]$\lambda5007$ and H$\beta$ line fluxes for the younger and older populations,
finding ratios of $0.3\pm0.1$ and $1.1\pm0.2$ respectively.  
The bluer galaxies lack a clear [OIII] feature, although the difference with the redder 
galaxies is only marginally significant.

\section{Discussion}
\label{sec:discussion}

We present stacked grism spectroscopy of a mass-limited sample of 171 quiescent galaxies at $1.4<z<2.2$ from 
the 3D-HST treasury program \citep{Brammer12}.  We demonstrate that we can resolve absorption features such 
as the G-band ($\lambda$4304$\mathrm{\AA}$), H$\beta (\lambda$4861$\mathrm{\AA}$), Mg I ($\lambda$5175$\mathrm{\AA}$), 
and Na I ($\lambda$5894$\mathrm{\AA}$) in the median stacks due to the high quality grism redshifts.
We further detect centrally-concentrated H$\beta$ and [OIII] in emission, indicating either residual central
star-formation \citep[e.g.,][]{Bezanson13} or an AGN.

The average L$_{\mathrm{[OIII]}}$ derived from the stacks of $0.4\times10^{7}$ L$_{\odot}$ 
is consistent with the results of \citet{Mendel13}, who
find that 70-80\% of recently-quenched galaxies in the SDSS have LINER-like active nuclei (L$_{\mathrm{[OIII]}}\lesssim10^{7}$ L$_{\odot}$).
Furthermore, the [OIII]$\lambda5007$/H$\beta$ ratio of 0.9 for a median mass of $10^{10.83}$ M$_{\odot}$ places the average
quiescent galaxy within the LINER region of the Mass-Excitation diagnostic diagram \citep{Juneau11}, albeit marginally consistent with the ``composite'' 
crossover region within the error bars.
From the average dust-corrected [OIII] ($\lambda$5007$\mathrm{\AA}$) luminosity,
we infer that quiescent galaxies may have a typical AGN luminosity of $L_{bol}=2\times10^{42}$ erg s$^{-1}$.
These results are consistent with the recent work of \citet{Olsen12}, who find 
that 70--100\% of massive quiescent galaxies at $1.5<z<2.5$ contain a low- or high-luminosity AGN.
The potential ubiquitous nature of AGNs in massive, quiescent $z\sim2$ galaxies provides observational support that
black hole accretion may be more effective at the high-mass end \citep[e.g.,][]{Kriek07}.

To understand the relationship between rest-frame color and our spectroscopic age measurements, 
we further stack the data in two rest-frame color bins.  We find that the 
bluest quiescent galaxies have a more prominent H$\beta$ absorption line, with 
best-fit ages of $0.9^{+0.2}_{-0.1}$ Gyr, confirming the idea that these galaxies are the most recently quenched.  
Furthermore, we find that redder quiescent galaxies (80\% of the population) have an older
age of $1.6^{+0.5}_{-0.4}$ Gyr.  Although previous spectroscopic studies have measured similarly 
old ages \citep[e.g.,][]{Kriek09a, Onodera12, vandeSande12}, this is the first time we are able to probe the full
population.  
These results suggest that age varies significantly with the colors of quiescent galaxies, consistent
with \citet{Whitaker10}.  

Galaxies with the reddest rest-frame colors likely have some degree of dust extinction, as dust-free models 
cannot produce such red colors at these redshifts.  Dust extinction does not greatly affect the 
line-index measurements for single stellar populations given
the narrow baseline of the spectral features, but its effect can be significant for the 4000\AA\ break \citep{MacArthur05}.
Consequently, although the absorption features of the 3D-HST stacks will be effectively independent of dust extinction,
the rest-frame color can change significantly in the presence of dust due to its sensitivity to the continuum shape.
We cannot rule out that the reddest galaxies have a potentially significant contribution from younger, dusty
quiescent galaxies.  Nonetheless, despite this caveat, we measure a significant age difference 
between the two populations of 0.7 Gyr that is consistent with the observed rest-frame color differences.

The key result from this paper is that we have conclusively demonstrated that
old, quiescent galaxies exist in large numbers at $z\sim2$. This is not a surprise given extensive 
previous results that were based on broadband and medium-band photometry \citep[e.g.,][]{Brammer11}, and 
ground-based spectroscopy for small samples \citep[e.g.,][]{Onodera12},
but our detection of metal lines in the stacked spectrum puts any lingering concerns to rest. 
Our result also implies that the ``UVJ technique'', first described in \citet{Labbe05}, is very effective 
in identifying quiescent galaxies at $z\sim2$.

\begin{acknowledgements}
We sincerely thank the referee for the thorough review of the manuscript.
This research was supported by an appointment to the NASA Postdoctoral Program at 
the Goddard Space Flight Center, administered by Oak 
Ridge Associated Universities through a contract with NASA.
Support from STScI grant GO-12177 and NASA ADAP grant NNX11AB08G 
is gratefully acknowledged.
\end{acknowledgements}

\addcontentsline{toc}{chapter}{\numberline {}{\sc References}}

\end{document}